\documentclass[12pt,thmsa]{article}
\usepackage{amsfonts}
\usepackage{graphicx}
\usepackage{epsfig}
\usepackage{graphics}

\makeatletter
\newcommand{\row}[1]{\mathord{\buildrel{\lower3pt\hbox{$\scriptscriptstyle\rightarrow$}}\over #1}}

\newcommand{\dyadic}[1]{\mathord{\dyadic@rrow{#1}}}
\newcommand{\dyadic@rrow}[1]{
\begin{picture}(12,12)(-1,0)
\put(-2,12){\makebox(0,0)[t]{$\scriptscriptstyle\downarrow$}}
\put(-2,12){\makebox(0,0)[l]{$\scriptscriptstyle\longrightarrow$}}
\put(5,0){\makebox(0,0)[b]{$#1$}}
\end{picture}
}
\newcommand{\bra}[1]{\bigl\langle #1 \bigr|}
\newcommand{\ket}[1]{\bigl| #1 \bigr\rangle}

\input{tcilatex}
\begin{document}

\begin{center}
\bigskip {\large \ Transfer  information remotely via noise  entangled coherent channels }

\bigskip A. El Allati$^{\dag}$, N. Metwally $^{\ddag}$ and Y. Hassouni$^{\dag}$\\[0pt]
$^{\dag}$ Laboratoire de Physique Théorique, Département de Physique
Faculté des sciences, Université Mohammed V - Agdal Av. Ibn
Battouta, B.P. 1014, Agdal, Rabat, Morocco \\
$^{\ddag}$Mathematics
Department, College of Science, University of Bahrain, 32038 Bahrain

\end{center}

\begin{abstract}
In this contribution, a generalized protocol of quantum
teleportation is suggested to  investigate the possibility of
remotely transfer unknown multiparities  entangled coherent state.
A theoretical technique is introduced to generate maximum
entangled coherent states  which are used as quantum channels. We
show that the mean photon number plays a central role on the
fidelity of the transferred information. The noise parameter can
be considered as a control parameter only for small values of the
mean photon number.
\end{abstract}

\section{Introduction}

Quantum information is one of the most important achievements of
this century, where one can overcome the problems of handling
confidential information. There are different ways to communicate
the quantum information. As an example, quantum teleportation,
which one can use it to send information remotely in a direct way
\cite{ben,yeo}, quantum coding, where the given information is
coded in a different state, which is sended to the receiver who
decodes the information \cite{ben1,mer}, quantum cryptography,
which is another technique to distribute quantum key between two
users to communicate safely \cite{ben2,Ekert}.

To achieve theses tasks, one needs entangled pairs, which
represent quantum channels between the sender and the receiver,
local operations and measurements. Since entangled pairs are
crucial resource in quantum communication, the preparation of
maximally entangled states is a crucial task. There are several
attempts carried out to generate entangled channels
between different types(see \cite%
{Wang,Juan,Lee,Shelly} for recent references).

One of the promising type of  entangled states are the coherent
states, which are used widely in the context of quantum
information. As an example, Zhou and Yang \cite{Ling} have used
them to transfer entanglement between atomic state and two-modes
cavity state. Also, they has been used to perform an optimal
teleportation \cite{Mari}. The coherent states are employed to
implement a probabilistic teleportation scheme, in which the
amount of classical information send by Alice is restricted to one
bit \cite{Bin}. Enk and Hirota \cite{Hiro}, have studied another
type of coherent states produced from Schr\"odinger cat states by
using a 50/50 beam splitter. This class of coherent states which
can be used to teleport one qubit where, a simple protocol that
achieves this aim with a $50\%$ probability of success is
introduced. Also, Wang has proposed a simple scheme to teleport
both the bipartite and multipartite by using only linear optical
devices such as beam splitters and phase shifters, and two-mode
photon number measurements \cite{Wang1}.

In reality, one can generate maximum entangled states, MES, but
keeping them isolated is impossible task. Therefore, these MES
turn into partially entangled states, PES due to the undesirable
interactions. So, there are great efforts have been done to
investigate the possibility of performing quantum information
tasks by using these partially entangled states. In this context,
Enk has considered the decoherence of multi-dimensional entangled
coherent states due to photon absorption losses  \cite{Enk} .

In this article, we introduce a general scheme of quantum
teleportation by using maximum and partial entangled states. Also,
this generalized protocol is employed to transfer quantum
information through noise quantum channels. This paper is
organized as follows: In Sec.$2$, we introduce a quantum
teleportation scheme to teleportate a tripartite entangled
coherent state by using a quantum channel consists of four parties
coherent state. This entangled channel is different from that has
been used in \cite{Wang1}. The generalization of this
teleportation protocol is the subject of Sec.$3$. Implementation
of quantum teleportation via partially entangled state is
described in Sec.$4$. Finally Sec.$5$, is devoted to discuss our
results.

\section{Teleportation scheme for tripartite state}

Entangled coherent states have been proposed as a potential
quantum channels to teleport unknown quantum states. These
coherent sates can be written as
function of the Fock state \cite{Gil}(photon number state) $\bigl| n %
\bigr\rangle$,
\begin{equation}
\bigl| \pm\alpha \bigr\rangle=\exp(-2|\alpha|^2)\sum_{n}^{\infty}{\frac{%
(\pm\alpha)^n}{\sqrt{n!}}\bigl| n \bigr\rangle}.
\end{equation}
An  entangled coherent state between two modes  can be written as,
\begin{equation}  \label{EC2}
\bigl| \alpha \bigr\rangle_{12}=(\bigl| \alpha,\alpha \bigr\rangle+\bigl| %
-\alpha,-\alpha \bigr\rangle)/N_{\phi}
\end{equation}
where $N_{\phi}=2-2exp[-4|\alpha|^2]$ and $\left\langle
\alpha|-\alpha \right\rangle=exp(-2(|\alpha|^2)$\cite {Wang1,Enk}.
This entangled  state (\ref{EC2}), has
been used by Enk and Hirota to teleport a  Schr\"{o}dinger cat state \cite%
{Enk}. Wang has used a tripartite entangled  coherent state of the form,
\begin{equation}
|\phi\rangle^{\pm}=N_{\alpha}^{\pm}(\bigl| \sqrt{2}\alpha,\alpha,\alpha %
\bigr\rangle_{123}\pm\bigl| -\sqrt{2}\alpha,-\alpha,-\alpha \bigr\rangle),
\end{equation}
where, $N_{\alpha}^{\pm}=[2(1\pm e^{-8|\alpha|^{2}})]^{-\frac{1}{2}}$ is the
normalized factor, to teleporte two qubits entangled coherent state \cite%
{Wang1}. Also, the same author has suggested a coherent state of
four particles to transfer a tripartite entangled coherent state.

In this current protocol, we introduce a different class  of
entangled coherent states  consists of  four qubit as a quantum
channel to teleport tripartite entangled coherent state. The
advantage of this choice is:  it can be generalized to
multipartite entangled channels easily, where we have generalized
our results theoretically to generate a family of maximum
entangled coherent states of mode $m$. This family can be used to
transport a family of $(m-1)$ modes of ECS by using the suggested
generalized scheme. In the following subsections we discuss these
phenomena by using maximum and partial entangled coherent states
as quantum channels.

\subsection{Using Maximum entangled as a quantum channel}

Consider that Alice is given a tripartite coherent state defined
as,
\begin{eqnarray}
\rho _{u} &=&|\kappa _{1}|^{2}\bigl|\sqrt{2}\alpha ,\alpha ,\alpha %
\bigr\rangle_{123}\bigl\langle\sqrt{2}\alpha ,\alpha ,\alpha \bigr|+\kappa
_{1}\kappa _{2}^{\ast }\bigl|\sqrt{2}\alpha ,\alpha ,\alpha \bigr\rangle%
_{123}\bigl\langle-\sqrt{2}\alpha ,-\alpha ,-\alpha \bigr|  \nonumber \\
&&+\kappa _{1}^{\ast }\kappa _{2}\bigl|-\sqrt{2}\alpha ,-\alpha ,-\alpha %
\bigr\rangle_{123}\bigl\langle\sqrt{2}\alpha ,\alpha ,\alpha \bigr|+|\kappa
_{2}|^{2}\bigl|-\sqrt{2}\alpha ,-\alpha ,-\alpha \bigr\rangle_{123}%
\bigl\langle-\sqrt{2}\alpha ,-\alpha ,-\alpha \bigr|.  \nonumber
\\
\end{eqnarray}%
 The aim of Alice is
sending this unknown state to Bob, who shares with her a maximum
multiparities entanglement coherent state (ECS),
\begin{eqnarray}  \label{channel}
\rho ^{\pm } &=&\frac{1}{N_{\pm }^{2}}\Bigl\{\bigl|2\alpha ,\sqrt{2}\alpha
,\alpha ,\alpha \bigr\rangle_{4567}\bigl\langle2\alpha ,\sqrt{2}\alpha
,\alpha ,\alpha \bigr|\pm \bigl|2\alpha ,\sqrt{2}\alpha ,\alpha ,\alpha %
\bigr\rangle_{4567}\bigl\langle-2\alpha ,-\sqrt{2}\alpha ,-\alpha ,-\alpha %
\bigr|  \nonumber \\
&\pm &\bigl|-2\alpha ,-\sqrt{2}\alpha ,-\alpha ,-\alpha \bigr\rangle_{4567}%
\bigl\langle2\alpha ,\sqrt{2}\alpha ,\alpha ,\alpha \bigr|  \nonumber \\
&+&\bigl|-2\alpha ,- \sqrt{2}\alpha ,-\alpha ,-\alpha \bigr\rangle_{4567}%
\bigl\langle-2\alpha ,-\sqrt{2}\alpha ,-\alpha ,-\alpha \bigr|\Bigr\},
\nonumber \\
&&
\end{eqnarray}%
where $N_{\pm }=\sqrt{2(1\pm e^{-16|\alpha |^{2}})}$ is the
normalization factor. This density operator represents two class
of states: maximum  and partially entangled states. It behaves as
MES between the system $4$ and the systems $5,6,7$, where the
concurrence \cite{Wootter}
in this case $\mathcal{C}_{4,(567)}^{-}=1$, while $\mathcal{C}%
_{4,(567)}^{+}=tanh(8|\alpha |^{2})$. On the other hand, the
density operator $\rho ^{- }$ behaves as a PES for the other
partitions, where
\begin{equation}
\mathcal{C}_{5(467)}^{\pm }=\frac{\sqrt{1-e^{-8|\alpha |^{2}}}\sqrt{%
1-e^{-24|\alpha |^{2}}}}{1\pm e^{-16|\alpha |^{2}}},\quad \mathcal{C^{\pm }}%
_{6(457)}=\frac{\sqrt{1-e^{-4|\alpha |^{2}}}\sqrt{1-e^{-28|\alpha |^{2}}}}{%
1\pm e^{-16|\alpha |^{2}}}=\mathcal{C}_{7(456)}^{\pm }
\end{equation}

Let us assume that, the partners use the maximum entangled state
$\rho^{-}$ as a quantum channel. Then the total state of the system is $%
\rho_u\otimes\rho_{4567}^{-}$. We can summarize the steps of
implementing the teleportation protocols as follows:

\begin{enumerate}
\item Alice mixes the unknown state $\rho_u$, with the quantum
channel $\rho_{4,567}^{-}$ by applying a
series of operations defined by the beam splitters and phase shifts \cite%
{Wang1, Rab}. In an explicit form
\begin{equation}
\rho_{u}\otimes\rho_{4567}^{-}\rightarrow
R_{34}R_{31}R_{32}\rho_{u}\otimes\rho_{4567}^{-}R_{32}^*R_{31}^*R_{34}^*=%
\rho_{out}
\end{equation}
where $R_{ij}\ket{ \mu }\ket{\nu}=\ket{ \frac{
\mu+\nu}{\sqrt{2}}}_i\ket{\frac{\mu-\nu}{\sqrt{2}}}_j$ . Theses
operations separate the systems $1$ and $2$ from the initial
system. So, the final output state $\rho_{out}$ is a direct
product of the vacuum
states $\bigl| 0 \bigr\rangle_1\bigl\langle 0 \bigr|\otimes\bigl| 0 %
\bigr\rangle_2\bigl\langle 0 \bigr|$ and the final unnormalized state,

\begin{eqnarray}
\rho_{f}&=&|\kappa_1|^2\bigl| \psi_1 \bigr\rangle\bigl\langle \psi_1 \bigr| %
-|\kappa_1|^2\bigl| \psi_1 \bigr\rangle\bigl\langle \psi_2 \bigr|%
-\kappa_1\kappa_2^*\bigl| \psi_1 \bigr\rangle\bigl\langle \psi_3 \bigr|+
\kappa_1\kappa_2^*\bigl| \psi_1 \bigr\rangle\bigl\langle \psi_4 \bigr|
\nonumber \\
&&+|\kappa_1|^2\bigl| \psi_2 \bigr\rangle\bigl\langle \psi_2 \bigr|%
+|\kappa_1|^2\kappa_2^*\bigl| \psi_2 \bigr\rangle\bigl\langle \psi_2 \bigr|%
-\kappa_1\kappa_2^*\bigl| \psi_2 \bigr\rangle\bigl\langle \psi_3 \bigr| %
+\kappa_1\bigl| \psi_2 \bigr\rangle\bigl\langle \psi_4 \bigr|  \nonumber \\
&&-\kappa_2\kappa_1^*\bigl| \psi_3 \bigr\rangle\bigl\langle \psi_1 \bigr|%
+\kappa_2\kappa_1^*\bigl| \psi_3 \bigr\rangle\bigl\langle \psi_2 \bigr|%
+|\kappa_3|^2\bigl| \psi_3 \bigr\rangle\bigl\langle \psi_3 \bigr| %
+|\kappa_3|^2\bigl| \psi_3 \bigr\rangle\bigl\langle \psi_4 \bigr|  \nonumber
\\
&&-\kappa_2\kappa_1^*\bigl| \psi_4 \bigr\rangle\bigl\langle \psi_1 \bigr|%
-\kappa_2\kappa_1^*\bigl| \psi_4 \bigr\rangle\bigl\langle \psi_2 \bigr|%
-|\kappa_2|^2\bigl| \psi_4 \bigr\rangle\bigl\langle \psi_3 \bigr| %
+|\kappa_2|^2\bigl| \psi_4 \bigr\rangle\bigl\langle \psi_4 \bigr|,
\end{eqnarray}
where,
\begin{eqnarray}
\ket{ \psi_1}&=&\ket{ 2\sqrt{2}\alpha,0,\sqrt{2}%
\alpha,\alpha,\alpha},\quad \ket{ \psi_2}=\ket{\sqrt{2}\alpha,-\sqrt{2}\alpha,-\alpha,-\alpha }\nonumber \\
\ket{ \psi_3}&=&\ket{0,-2\sqrt{2}\alpha,\sqrt{2}%
\alpha,\alpha,\alpha},\quad \ket{\psi_4 \bigr\rangle=\bigl| -2
\sqrt{2}\alpha,0,-\sqrt{2}\alpha,-\alpha,-\alpha}.
\end{eqnarray}

\item Alice performs two photon number measurements on modes $3$ and $4$.
The probability to find $l$ and $n$ photon, $P(l,n)$ in modes $3$ and $4$ is
given by
\begin{equation}  \label{prob}
\mathcal{P}(l,n)=\Bigl|\bigl\langle l,n \bigr|\rho_{f}\bigl| l,n \bigr\rangle%
\Big|^2.
\end{equation}
The probability, $\mathcal{P}(l,n)=0$, if both $l$ and $n$ are non
zero. However, if $n\neq0$ and $l=0$, the state on Bob'side
collapses into,
\begin{eqnarray}  \label{bob1}
\rho^{n}_{Bob}&=&\lambda_1\ket{\sqrt{2}\alpha,\alpha,\alpha}\bra{\sqrt{2}\alpha,\alpha,\alpha}
-\lambda_2 \ket{\sqrt{2}
\alpha,\alpha,\alpha}\bra{-\sqrt{2}\alpha,-\alpha,-
\alpha} \nonumber \\
&-&\lambda_3\ket{-\sqrt{2}\alpha,-\alpha,-\alpha}\bra{
\sqrt{2}\alpha,\alpha,\alpha} +\lambda_4\ket{-\sqrt{2}%
\alpha,-\alpha,-\alpha}\bra{-\sqrt{2}\alpha,-\alpha,-\alpha}
\end{eqnarray}
where $\lambda^{(n)}_1=\frac{|\kappa_{1}|^2}{N_1}, \lambda^{(n)}_2=%
\frac{\kappa_{1}\kappa_{2}^*}{N_1}(-1)^n, \lambda^{(n)}_3= \frac{
\kappa_{1}^*\kappa_{2}}{N_1}(-1)^n, \lambda^{(n)}_4=\frac{%
|\kappa_{2}|^2}{N_1}(-1)^{2n}$ and $N_1=|\kappa_{1}|^{2}+|%
\kappa_{2}|^{2}-2(-1)^{n}e^{-8|\alpha|^{2}}Re(\kappa_{2}^{\ast}%
\kappa_{1})$ is the normalized factor.

\item Alice sends her results through classical channel to Bob,
who performs a three $\pi$ phase shifters of modes $5, 6$ and $7$
local transformation on his state $\rho_{Bob}$, i.e., Bob applies
the unitary operator,
\begin{equation}
U_{P}=
e^{-i\pi(a_{5}^{\dagger}a_{5}+a^{\dagger}_{6}a_{6}+a^{\dagger}_{7}a_{7})}
\end{equation}
If the integer $n$ is odd, then  Bob gets exactly the same input state $\rho_{u}$%
. However if  $n$ is even, Bob performs an extra operation  such
that,
\begin{eqnarray}
\ket{\sqrt{2}\alpha,\alpha,\alpha}\rightarrow \ket{\sqrt{2}%
\alpha,\alpha,\alpha}, \quad
\ket{-\sqrt{2}\alpha,-\alpha,-\alpha}\rightarrow& -\ket{-\sqrt{2}%
\alpha,-\alpha,-\alpha},
\end{eqnarray}
to get the teleported state exactly.

Now, if we assume that $n$ is an odd integer and $l=0$, then the probability
of successes (\ref{prob}) is given by
\begin{equation}
\mathcal{P}(n,0)=\frac{e^{-8|\alpha|^2}|2\sqrt{2}\alpha|^{2n}}{%
2n!(1-e^{-16|\alpha|^2)},}
\end{equation}
which is independent of the parameter $\kappa_{1,2}$. This
probability becomes $0.5$, when $|\alpha|\rightarrow \infty$.

Let us consider the case, $n=0$ and $n\neq0$. In this case the
state of Bob collapses into,

\begin{eqnarray}
\rho_{Bob}^{(l)}&=&\lambda^{(l)}_1\bigl| -\sqrt{2}\alpha,-\alpha,-\alpha %
\bigr\rangle\bigl\langle {-\sqrt{2}\alpha,-\alpha,-\alpha} \bigr| %
-\lambda^{(l)}_2\bigl| -\sqrt{2}\alpha,-\alpha,-\alpha \bigr\rangle%
\bigl\langle {\sqrt{2}\alpha,\alpha,\alpha} \bigr|  \nonumber \\
&-& \lambda^{(l)}_3\bigl| \sqrt{2}\alpha,\alpha,\alpha \bigr\rangle%
\bigl\langle {-\sqrt{2}\alpha,-\alpha,-\alpha} \bigr| +\lambda^{(l)}_4\bigl|
\sqrt{2}\alpha,\alpha,\alpha \bigr\rangle\bigl\langle \sqrt{2}%
\alpha,\alpha,\alpha \bigr|,
\end{eqnarray}

where, $\lambda^{(l)}_1=\frac{|\kappa_{1}|^2}{N_2}, \lambda^{(l)}_2=\frac{%
\kappa_{1}\kappa_{2}^*}{N_2}(-1)^l, \lambda^{(l)}_3=\frac{%
\kappa_{1}^*\kappa_{2}}{N_2}(-1)^l, \lambda^{(l)}_4=\frac{%
|\kappa_{2}|^2}{N_1}(-1)^{l}$ and $N_2=|\kappa_{1}|^{2}+|%
\kappa_{2}|^{2}-2(-1)^{l}e^{-8|\alpha|^{2}}Re(\kappa_{2}^{\ast}%
\kappa_{1})$ is the normalized factor. Also, in this case the
probability of successes $\mathcal{P}(0,l)=\mathcal{P}(n,0)$.
\end{enumerate}

\subsection{Using Partially entangled state as a quantum channel}

In sec.(2.1), we show that the density operator sometimes behaves
as a non maximum entangled state, i.e., PES. Let us assume that
Alice has the possibility of using only this class of states as
quantum channels to teleportate a tripartite entangled. In this
case the total state of the system is
$\rho_{u}\otimes\rho^{+}_{4567}$. The partners perform the
steps(1-3) as described in Sec.(2.1). If Alice measures $n$
photons in the mode $3$ and $(l=0)$ photons in mode $4$, then the
partners  end  the protocol with a similar state as (\ref{bob1}).
The only difference between them is the sign of the second and
third terms are positive. To achieve this protocol with fidelity
$1$, one consider $n$ is an even number. In this case the
probability of successes is given by,
\begin{equation}
\mathcal{P}=\frac{(1-e^{-8|\alpha|^2})^2}{2(1+e^{-16|\alpha|^2})}.
\end{equation}
It is clear that $\mathcal{P}$ depends on the parameter $\alpha$,
but is
independent of the parameters $\kappa_{1,2}$. In the limit $%
|\alpha|\rightarrow \infty$, the probability of success becomes
$\frac{1}{2}$ (one ebit) and $\mathcal{P}<\frac{1}{2}$, in this
case quantum channel is not a MES.\

\section{Generalized quantum teleportation Protocol}

In this section, we generalized the results of Sec.$2$. For this
aim, we define a non orthogonal maximum entangled as
\begin{eqnarray}  \label{gen}
|\Psi\rangle^{\pm}_{0..m} &=& A^{\pm}_{m+1}\Bigl(|2^{\frac{m-1}{2}%
}\alpha\rangle_{0}...|2^{\frac{1}{2}}\alpha\rangle_{m-2}|\alpha%
\rangle_{m-1}|\alpha\rangle_{m}  \nonumber \\
&&\pm|-2^{\frac{m-1}{2}}\alpha\rangle_{0}...|-2^{\frac{1}{2}%
}\alpha\rangle_{m-2}|-\alpha\rangle_{m-1}|-\alpha\rangle_{m}|\Bigr),
\end{eqnarray}
where $A^{\pm}_{m+1}=[2(1\pm
e^{-2^{m+1}|\alpha|^{2}})]^{-\frac{1}{2}}$, is the normalized
factor. This class of states, which represent a quantum channel of
$m+1$ modes, can be used to teleporte a multiparities state of $m$
modes. To construct this maximum entangled multipartite state, one
can use the following table,
\begin{table}[htp!]
\begin{center}
\begin{tabular}{|c|c|c|c|c|c|c|c|}
\hline
$m$ &  &  &  &  &  &  &  \\ \hline
$0$ & $1$ &  &  &  &  &  &  \\
$1$ & 1 & 1 &  &  &  &  &  \\
$2$ & $\sqrt{2}$ & $1$ & $1$ &  &  &  &  \\
$3$ & $2$ & $2\sqrt{2}$ & $1$ & $1$ &  &  &  \\
$4$ & $2\sqrt{2}$ & $2$ & $\sqrt{2}$ & $1$ & $1$ &  &  \\
$\vdots$ & $\vdots$ & $\vdots$ & $\vdots$ & $\vdots$ & $\vdots$ & $\vdots$ &
$\vdots$ \\
$\vdots$ & $\vdots$ & $\vdots$ & $\vdots$ & $\vdots$ & $\vdots$ & $\vdots$ &
$\vdots$ \\
$m+1$ & $2^{\frac{m-1}{2}}$ & $\cdots$ & $\cdots$ & $\cdots$ & $\cdots$ & $%
\cdots$ & $\cdots$ \\ \hline
\end{tabular}%
\end{center}
\caption{This table represents a simple scheme for generating a maximum
entangled state of $m$ modes.}
\end{table}
as an example for $m=4$, the state vector is given by,
\begin{equation}
\bigl| 2\alpha,\sqrt{2}\alpha,\alpha,\alpha \bigr\rangle_{3210}=%
\prod^{1}_{3-m}\ket{ 2^{\frac{m-1}{2}}\alpha}_m\ket{\alpha}_{0}
\end{equation}
To evaluate the degree of entanglement contained in the general
density operator $\rho_{gen}=\ket{ \psi^{\pm}}\bra{\psi^{\pm}}$
, where the sate vector $\ket{\psi^{\pm}}$ is given by(%
\ref{gen}), we evaluate the concurrence. For this aim, we divided the $m+1$%
-modes systems into two systems: the first, $A$ is system $0$ and
second $B$ is the all remaining system of modes $m$, where each
system is linearly independent with respect to $\alpha$ and
$-\alpha$ which spanning a two dimensional subspaces of the
Hilbert space. Consider a set of  the orthonormal basis
$\{|0\rangle_{x},|1\rangle_{x}\}$, $x=A,B$ which satisfy the
Gram-Schmidt theorem. These basis could be written in the basis of
the coherent state as following

\begin{eqnarray}
\bigl| 0 \bigr\rangle_{A}&=& \bigl| 2^{\frac{m-1}{2}}\alpha
\bigr\rangle_{0}
\nonumber \\
\bigl| 1 \bigr\rangle_{A}&=& \frac{\bigl| -2^{\frac{m-1}{2}}\alpha %
\bigr\rangle_{0}-_{0}\bigr\langle 2^{\frac{m-1}{2}}\alpha  \bigl|
{-2^{\frac{m-1}{2}}}\alpha
\bigr \rangle_{0} \bigl| 2^{\frac{m-1}{2}}\alpha \bigr\rangle_{0}}{\sqrt{%
1-(_{0}\langle2^{\frac{m-1}{2}}\alpha\bigl|
-2^{\frac{m-1}{2}}\alpha\rangle_{0})^{2}}}
\nonumber \\
\bigl| 0 \bigr\rangle_{B}&=& \bigl| 2^{\frac{m-2}{2}}\alpha \bigr\rangle%
_{1}\bigl| 2^{\frac{m-3}{2}}\alpha \bigr\rangle_{2}...\bigl| \alpha %
\bigr\rangle_{m}  \nonumber \\
\bigl| 1 \bigr\rangle_{B}&=&\frac{\bigl| -2^{\frac{m-2}{2}}\alpha %
\bigr\rangle_{1}..\bigl|- \alpha\bigr\rangle_{m}-_{m}\bigr\langle\alpha\bigl|.._{1}\bigr\langle2^{%
\frac{m-2}{2}}\alpha\bigl|-
2^{\frac{m-2}{2}}\alpha\bigr\rangle_{1}..\bigl|-
\alpha\bigr\rangle_{m}|2^{\frac{m-2}{2}}\alpha\bigr\rangle_{1}..\bigl|\alpha\bigr\rangle_{m}} {%
\sqrt{1-(_{1}\bigr\langle2^{\frac{m-2}{2}}\alpha\bigl|.._{m}\bigr\langle\alpha\bigl|-\alpha%
\bigr\rangle_{m}..\bigl|-2^{\frac{m-2}{2}}\alpha\bigr\rangle_{1})^{2}}}
\end{eqnarray}

By using these new basis, state vector $\bigl| \psi
\bigr\rangle^{-}$ takes the form,
\begin{equation}
\bigl| \psi \bigr\rangle^{-}=x_{00}\bigl| 0 \bigr\rangle_{A}\bigl| 0 %
\bigr\rangle_{B}+x_{01}\bigl| 0 \bigr\rangle_{A}\bigl| 1 \bigr\rangle%
_{B}+x_{10}\bigl| 1 \bigr\rangle_{A}\bigl| 0 \bigr\rangle_{B}+x_{11}\bigl| 1 %
\bigr\rangle_{A}\bigl| 1 \bigr\rangle_{B},
\end{equation}
where,
\begin{eqnarray*}
x_{00} &=& A_{m}(1-e^{-2^{m+1}|\alpha|^{2}}), \\
x_{01} &=& - A_{m} e^{-2^{m}|\alpha|^{2}}\sqrt{1-e^{-2^{m+1}|\alpha|^{2}}},
\\
x_{10} &=& - A_{m} e^{-2^{m}|\alpha|^{2}}\sqrt{1-e^{-2^{m+1}|\alpha|^{2}}},
\\
x_{11} &=& -A_{m} \sqrt{1-e^{-2^{m+1}|\alpha|^{2}}}\sqrt{1-e^{-2^{m+1}|%
\alpha|^{2}}},
\end{eqnarray*}
 For this density operator the concurrence, $\mathcal{C}=1$. It
is clear that the degree of entanglement is independent of the parameters $%
\alpha$ and $m$.

Now, we  use the generalized density operator
$\rho^{-}=\ket{\psi^-}\bra{\psi^-}$, which is defined by the
generalized state vector (\ref{gen}) with $m+1$ modes to transfer
$m$-modes entangled coherent state. The partners, Alice and Bob
apply the protocol which is described in the previous section and
end the  protocol with a probability of successes,
\begin{equation}  \label{pn_o}
\mathcal{P}_{n_-odd}= \frac{e^{-2^{m}|\alpha|^{2}}|2^\frac{m}{2}\alpha|^{2n}}{%
2n!(1-e^{-2^{m+1}|\alpha|^{2}})}
\end{equation}
for odd $n$.  However if  $n$ is even, then the probability of
successes is,

\begin{equation}
\mathcal{P}_{n-even}=\frac{(1-e^{-2^{m}|\alpha |^{2}})}{2(1+e^{-2^{m+1}|%
\alpha |^{2}})}  \label{pn_e}
\end{equation}%
From Eqs.(\ref{pn_o}) and (\ref{pn_e}), the probability of
successes depends on the both of $\alpha $ and $m$ and
$\mathcal{P}\rightarrow 0.5$ as $\alpha \rightarrow \infty $ or
$m\rightarrow \infty $.
\section{Teleportation in the presences of noise}

The realistic investigation of quantum systems for quantum
information processing must take into account the decoherence
effect \cite{Hong}. The dynamical properties of coherent states in
the presences of noise have received considerable attention
\cite{Enk,Hyu}. There are several ways that the noise affects the
quantum channels, which may be used for the purposes of quantum
information tasks. Among these method is decoherence due to the
energy loss or the photon absorption \cite{Hiro,Enk,Hof}.

Assume that we have a source supplies the partners, Alice and Bob
with  maximum entangled coherent states. These entangled coherent
states propagate from the source to the locations of the partners.
Due to the interactions with the environment, the maximum
entangled coherent states turn into  partial entangled states,
where its degree of entanglement depends on the strength of the
noise. Consider that the source produces MECS defined by the
density operator, $\rho^{-}$ given by Eq.(\ref{channel}). The
effect of the noise is,

\begin{equation}
\rho_{PE}=U_{AE}\otimes U_{BE}\rho^{-}U^{\dagger}_{BE}\otimes
U^{\dagger}_{AE},
\end{equation}
where $U_{IE}\ket{\alpha}\ket{
0}_E=\ket{\sqrt{\eta}\alpha}_I\ket{\sqrt{1-\eta}\alpha}_E, I=A$,or
$B$ and $\ket{ 0}_E$ referees to the environment state. This
effect is equivalent to employing a half mirror for the noise
channel
\cite{Hiro}. In an explicit form, one can write the output density operator $%
\rho_{PE}$ as

\begin{eqnarray}  \label{Par}
\rho_{PE}&=& \frac{1}{{N_{\alpha}}}\Bigl[\bigl| 2\sqrt{\eta}\alpha,%
\sqrt{2}\sqrt{\eta}\alpha,\sqrt{\eta}\alpha,\sqrt{\eta}\alpha \bigr\rangle%
\bigl\langle 2\sqrt{\eta}\alpha,\sqrt{2}\sqrt{\eta}\alpha,\sqrt{\eta}\alpha,%
\sqrt{\eta}\alpha \bigr|  \nonumber \\
&+&\bigl| -2\sqrt{\eta}\alpha,-\sqrt{2}\sqrt{\eta}\alpha,-\sqrt{\eta}\alpha,-%
\sqrt{\eta}\alpha \bigr\rangle\bigl\langle -2\sqrt{\eta}\alpha,-\sqrt{2}%
\sqrt{\eta}\alpha,-\sqrt{\eta}\alpha,-\sqrt{\eta}\alpha \bigr|  \nonumber \\
&-&e^{-8|\alpha|^{2}}\bigl| 2\sqrt{\eta}\alpha,\sqrt{2}\sqrt{\eta}\alpha,%
\sqrt{\eta}\alpha,\sqrt{\eta}\alpha \bigr\rangle\bigl\langle -2\sqrt{\eta}%
\alpha,-\sqrt{2}\sqrt{\eta}\alpha,-\sqrt{\eta}\alpha,-\sqrt{\eta}\alpha %
\bigr|  \nonumber \\
&-& e^{-8|\alpha|^{2}}\bigl| -2\sqrt{\eta}\alpha,-\sqrt{2}\sqrt{\eta}\alpha,-%
\sqrt{\eta}\alpha,-\sqrt{\eta}\alpha \bigr\rangle\bigl\langle 2\sqrt{\eta}%
\alpha,\sqrt{2}\sqrt{\eta}\alpha,\sqrt{\eta}\alpha,\sqrt{\eta}\alpha \bigr|%
\Bigr],
\end{eqnarray}
where $N_{\alpha}=2(1-e^{-16|\alpha|^{2}})$ is the normalized
factor. To investigate how much the two states $\rho^{-}$ (the
input state) and the output state $\rho_{PE}$ are related to each
other, we evaluate the fidelity $\mathcal{F}$
\begin{figure}[b!]
\begin{center}
\includegraphics[width=18pc,height=12pc]{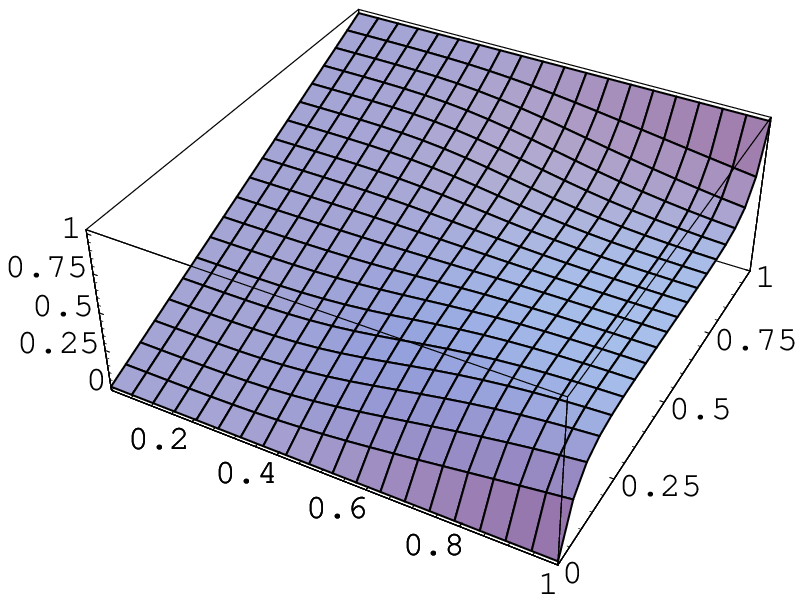}~
\includegraphics[width=18pc,height=12pc]{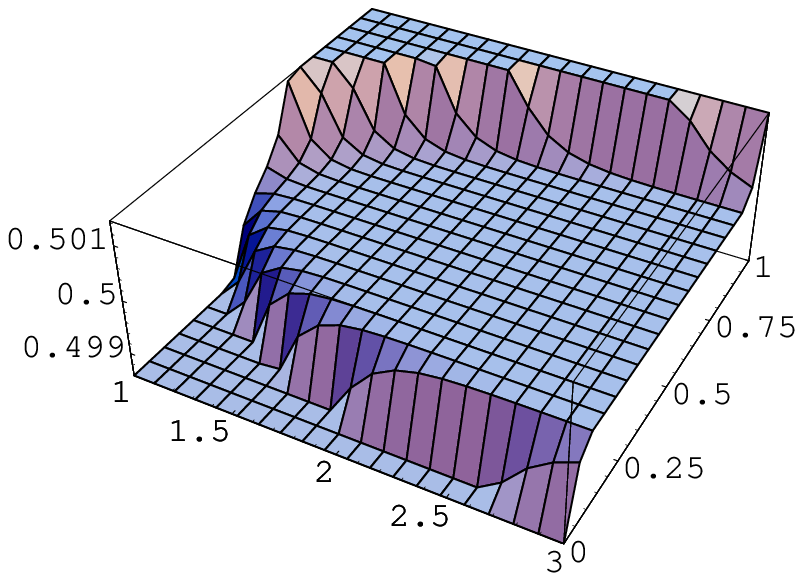}
\put(-20,40){$\eta$} \put(-250,40){$\eta$} \put(-130,20){$\alpha$}
\put(-370,15){$\alpha$} \put(-400,120){$(a)$}
\put(-175,120){$(b)$} \put(-450,70){$F$}
 \put(-210,70){$F$}
\end{center}
\caption{The fidelity of the input state as a function of $\alpha$
and $\eta$.}
\end{figure}
\begin{equation}
F=tr\{\rho^{-}\rho_{PE}\}= \frac{(1-e^{-16\eta|%
\alpha|^{2}})(1+e^{-16(1-\eta)|\alpha|^{2}})}{2(1-e^{-16|\alpha|^{2}})}
\end{equation}
\begin{figure}[t!]
\begin{center}
\includegraphics[width=18pc,height=12pc]{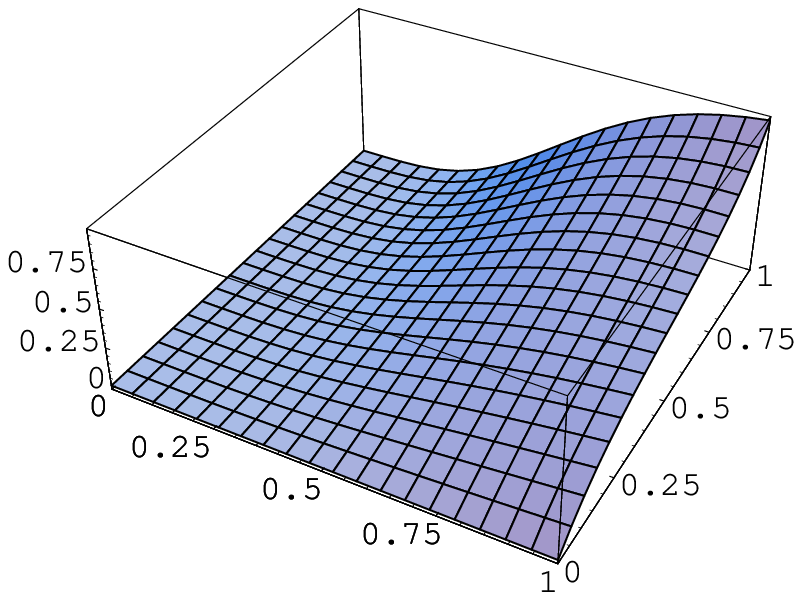}~
\includegraphics[width=18pc,height=12pc]{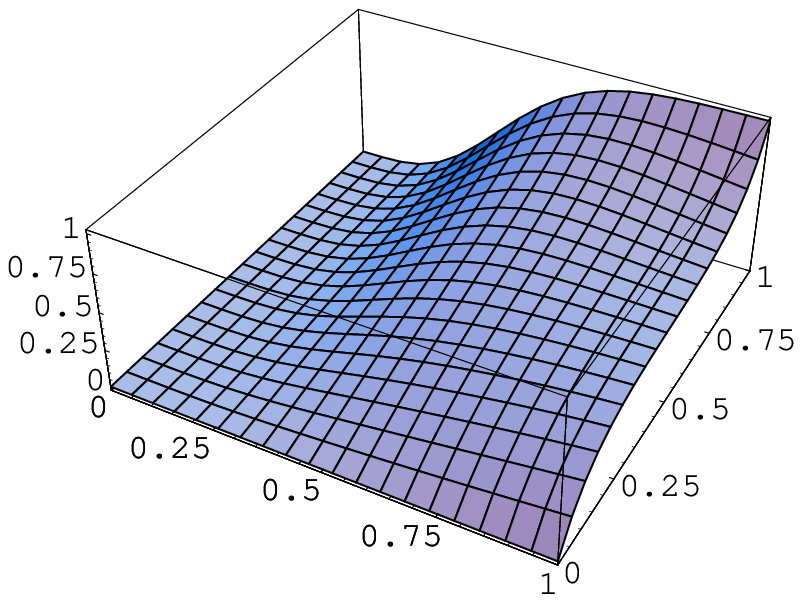}
\put(-20,40){$\eta$} \put(-250,40){$\eta$} \put(-150,15){$\alpha$}
\put(-370,15){$\alpha$} \put(-400,120){$(a)$}
\put(-175,120){$(b)$} \put(-450,70){$\mathcal{F}$}
 \put(-220,65){$\mathcal{F}$}\\
\includegraphics[width=15pc,height=15pc]{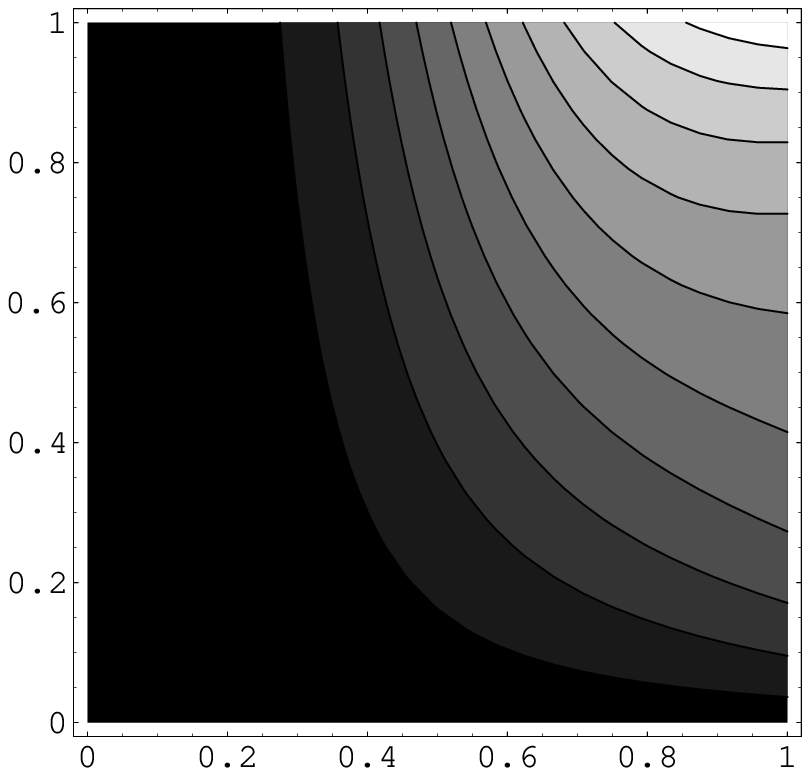}~\quad
\includegraphics[width=15pc,height=15pc]{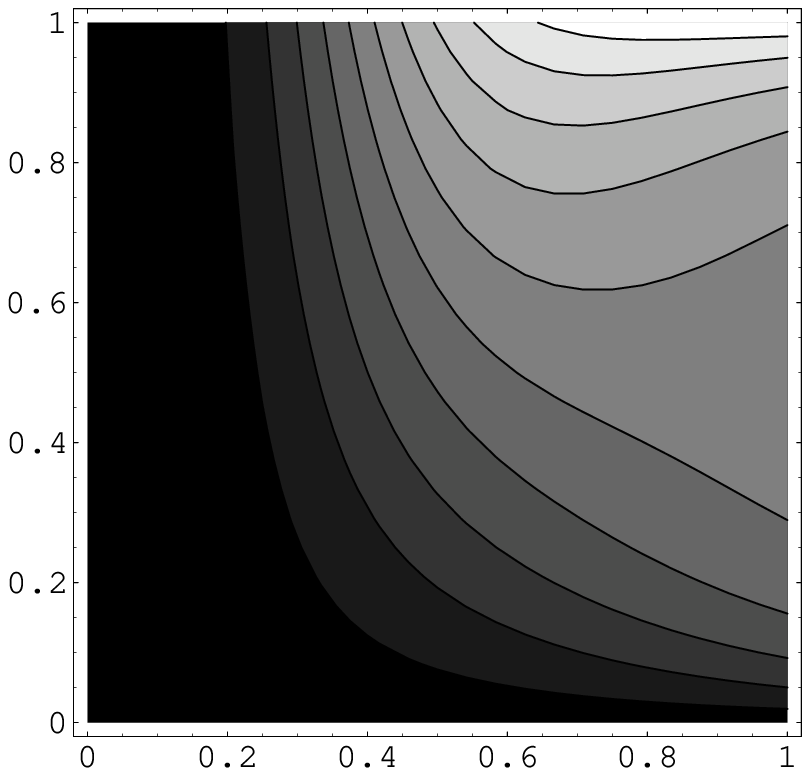}
\put(-180,100){$\eta$} \put(-380,100){$\eta$}
\put(-100,0){$\alpha$} \put(-280,0){$\alpha$}
\put(-370,200){$(c)$} \put(-140,200){$(d)$}
 \end{center}
\caption{The fidelity of the teleported state for  (a) $m=2$,
(b)$m=3$.}
\end{figure}
\begin{figure}[t!]
\begin{center}
\includegraphics[width=18pc,height=12pc]{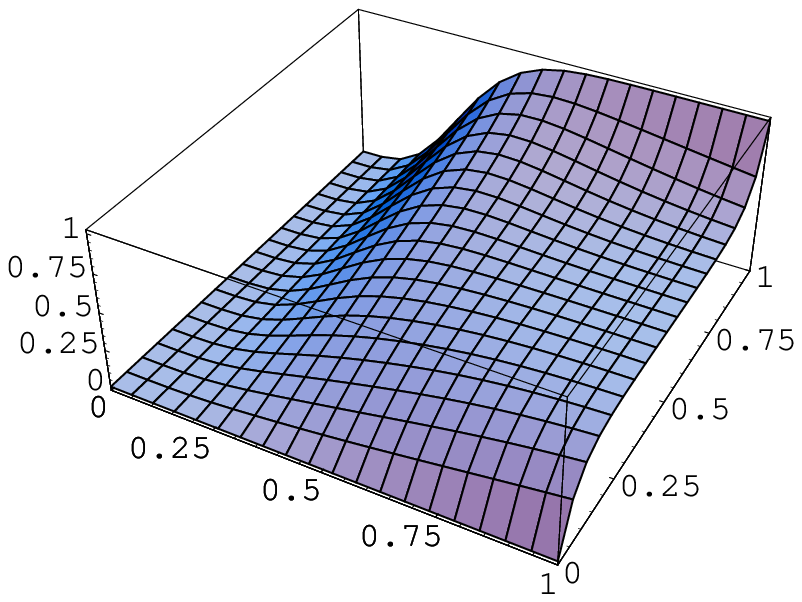}~
\includegraphics[width=18pc,height=12pc]{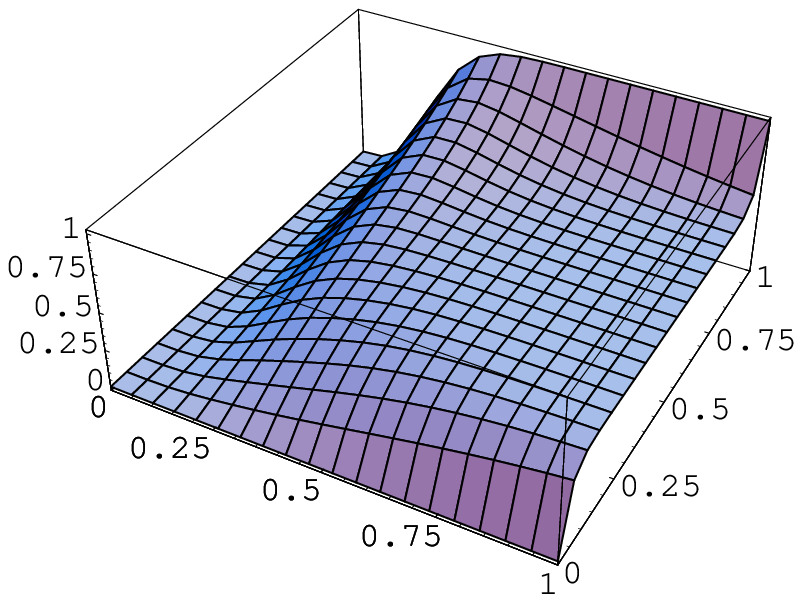}
\put(-20,40){$\eta$} \put(-250,40){$\eta$} \put(-150,15){$\alpha$}
\put(-370,15){$\alpha$} \put(-400,120){$(a)$}
\put(-175,120){$(b)$} \put(-450,70){$\mathcal{F}$}
 \put(-220,65){$\mathcal{F}$}\\
\includegraphics[width=15pc,height=15pc]{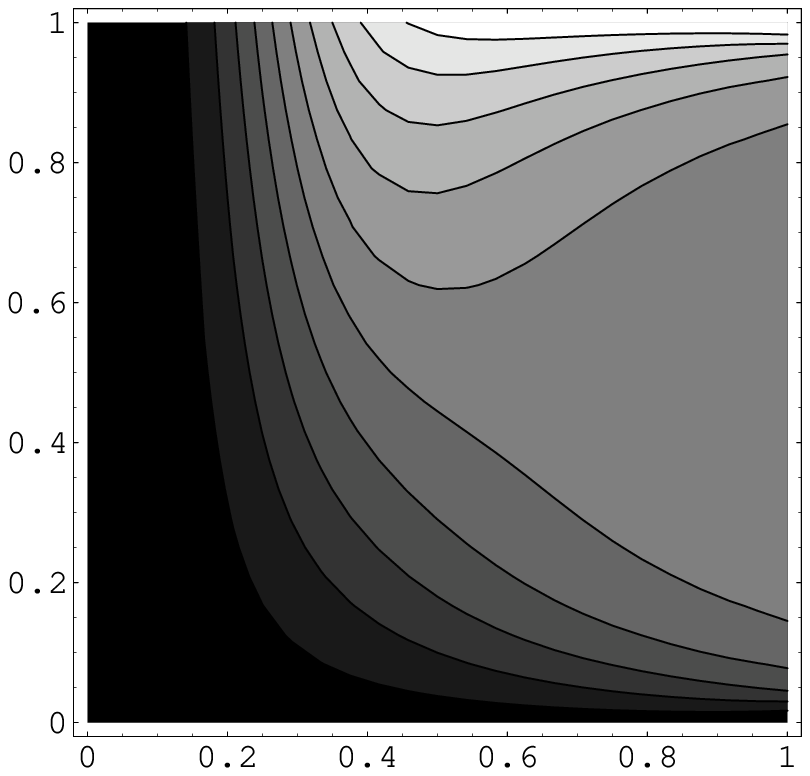}~\quad
\includegraphics[width=15pc,height=15pc]{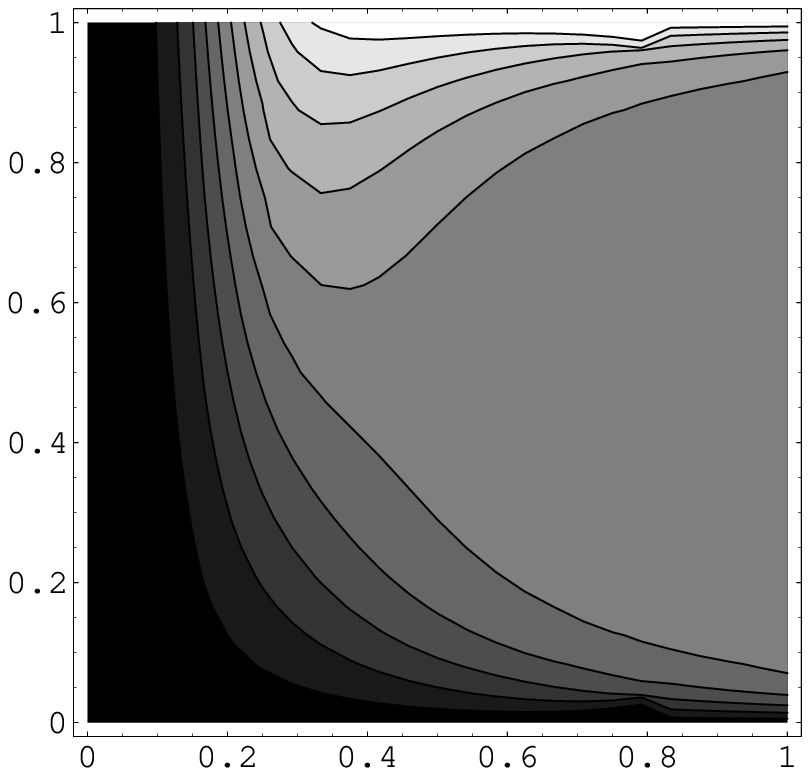}
\put(-180,100){$\eta$} \put(-380,100){$\eta$}
\put(-100,0){$\alpha$} \put(-280,0){$\alpha$}
\put(-370,200){$(c)$} \put(-140,200){$(d)$}
 \end{center}
\caption{The fidelity of the teleported state for  where $m=3$,
for Figs.$(a\&c)$ and $m=5$ for Fig.$(b\&d)$.}
\end{figure}
In Fig.(1), we plot the fidelity $F$, as a function of the
parameter $\alpha$ and the noise strength $\eta$. Fig.(1a),
displays the dynamic of the
fidelity, $F$ for small  value of $\alpha\in ]0,1]$, and $%
0\leq\eta\leq 1$. It is clear that for small values of $\eta\leq
0.5$ i.e, the correlation between the noise and the ECS is vary
strong, the fidelity of the output state is very small. However
$F$, increases as the noise strength $\eta$ increases, which is
means that the correlation between the coherent state and the
environmental noise is weak. Fig.(1b) shows the behavior of  the
fidelity for different range of $\alpha\in[1,3]$. In this case, we
have two different effective range of the noise strength $\eta$:
The first, when $0\leq \eta\leq 0.5$,  the fidelity increases and
reach its maximum value ($F=0.5$). The second one, for $0.5\le
\eta\leq 1$, the fidelity decrease  and reaches its minimum value
$(F=0.5)$. From these figures, one can say that, the effect of the
noise for small values of $\alpha$ is very small, but for a larger
values of $\alpha$, the noise effect is almost constant.

Let us assume that the quantum channel suffering from environmental noise.
So, the partners, Alice and Bob are forced to use the noise channel (\ref{Par}%
) to implement the quantum teleportation. If the given unknown state is of $%
m $ modes coherent state. The partners  use $(m+1)$ modes coherent
state as a quantum channel. They  apply the steps which are
described in the previous section and the final state at Bob's
hand is given by
\begin{equation}
\ket{\phi}_{m+1..2m}=(\kappa_{1}\ket{2^{\frac{m-1}{2}}
\alpha}......\ket{\alpha}_{m+1..2m}-(-1)^{n}
\kappa_{1}\ket{-2^{\frac{m-1}{2}}\alpha}.....\ket{-\alpha}_{m+1..2m})/\sqrt{N_{1}},
\end{equation}
where,
$N_{1}=|\kappa_{1}|^{2}+|\kappa_{2}|^{2}-(-1)^{n}e^{2^{m+1}|\alpha|^{2}}Re(\kappa_{2}^{*}\kappa_{1})$
 is the normalized factor. The fidelity of this state is,
\begin{equation}
\mathcal{F}_{m}=\frac{(1+e^{-2^{m}\eta ^{\prime 2}})(1-e^{-2^{m}\eta |\alpha |^{2}})}{%
2(1-e^{-2^{m}|\alpha |^{2}})},\quad
\end{equation}%
where $\eta ^{\prime }=1-\eta $, and $m\geq 1$.

 Fig.(2), shows  the  behavior of the fidelity  $\mathcal{F}(\alpha,\eta)$ for   different classes
 of the teleported state . This behavior
 is displayed for  small range of $\alpha$ and $0\leq\eta\leq 1$.
 In Fig.(2a), we set $m=2$, i.e., the teleported state is a
 bipartite entangled coherent state, while the used quantum channel
 is a tripartite ECS. This figure shows, that for small values of
 $\alpha<0.5$, i.e., the mean photon number is very small, the
 tripartite ECS, can not teleport the bipartite ESC, even for
 large value of $\eta$. However as the mean photon number
 increases the fidelity $\mathcal{F}$ increases and reaches its
 maximum value $(>0.75)$ at $\eta=1$. In Fig.(2b), we investigated
 the propagation of $\mathcal{F}$, for $m=3$, i.e., Alice, is
 asked to teleport a tripartite ECS  to Bob  via  a four ECS as a
 quantum channel. In this case the  intervals in which the quantum
 channel fails to teleport the required state is smaller than that
 depicted in Fig.(2a). On the other hand, $\mathcal{F}$, increases
 gradually as $\eta$ increases and for $\alpha>0.75$, the fidelity
 of the teleport state is almost unity at $\eta=1$.

 These phenomena are clearly shown in Figs.(1c$\&$ 2d), where we plot
 the fidelity $\mathcal{F}$ as a contour. It is clear that for
 small value of $\eta$, the fidelity is almost zero which
 appears as a dark region. However, as one increases the noise
 strength, the fidelity increases and reaches its maximum value at
 $\eta=1$. This behavior is shown in the bright region. The
 brightness (high fidelity) and darkness (low fidelity) regions can
 be determined simply from the figures, where the brightness appears
 for $\eta\geq 0.5$ for $m=2$ (Fig.(2c)), while for $m=3$, the
 brightness appears at $\eta\geq 0.29$. Also, the degree of
 brightness indicates the degree of fidelity.

The fidelity of a different class of ECS is displayed in Fig.(3a),
where we consider $m=4$. In this case, we can notice three
different behaviors of the fidelity, $\mathcal{F}$.  The first is
$\mathcal{F}=0$, for smaller values of $\alpha$. The second, for
$0\leq \eta<0.5.$, the fidelity increases to reach its maximum
value $(0.5)$, while for $0.5\leq\eta\leq 1$, $\mathcal{F}$
decrease  gradually to reach its minimum value(0.5). The same
behavior is depicted in Fig.(3b), where we consider $m=5$.

Also, In Fig.(3c), the fidelity $\mathcal{F}(\alpha,\eta)$ is
plotted as a contour graph. It is shown  as $m$ increases the
ability of transforming information is much better, where the dark
region appears only for small range of the noise strength $\eta$.
This appears clearly by comparing Fig.(1c), where $m=2$ and
Fig.(3d), where $m=5$.

For larger interval of the mean photon number $\alpha$, the noise
effect is very small comparing with that for small values of
$\alpha$. Theses results are shown in Fig.(4a) and Fig.(4b), where
different classes of the teleported states are considered ($m=3$
and $m=5$) respectively. From theses figures it is clear that the
maximum value of $\mathcal{F}=0.55$ for tripartite ECS. As one
increases $m$, the maximum value is almost $0.5$.
\begin{figure}
\begin{center}
\includegraphics[width=18pc,height=12pc]{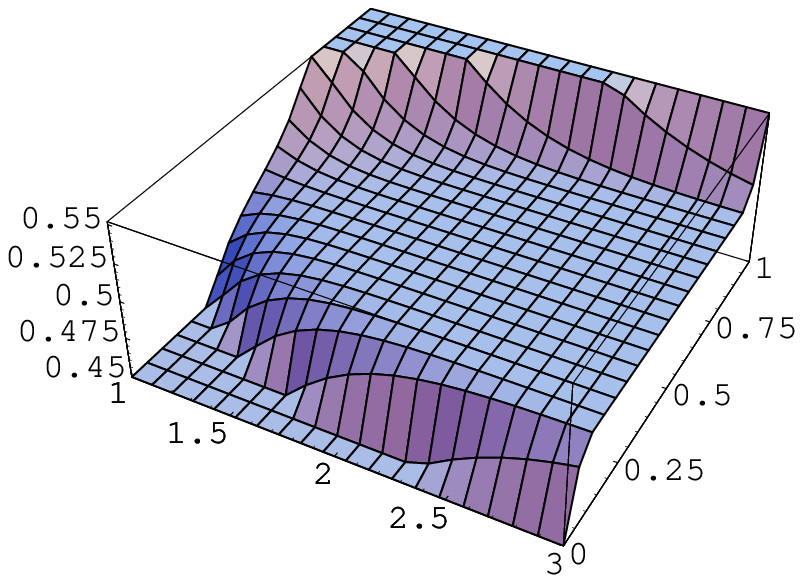}~
\includegraphics[width=18pc,height=12pc]{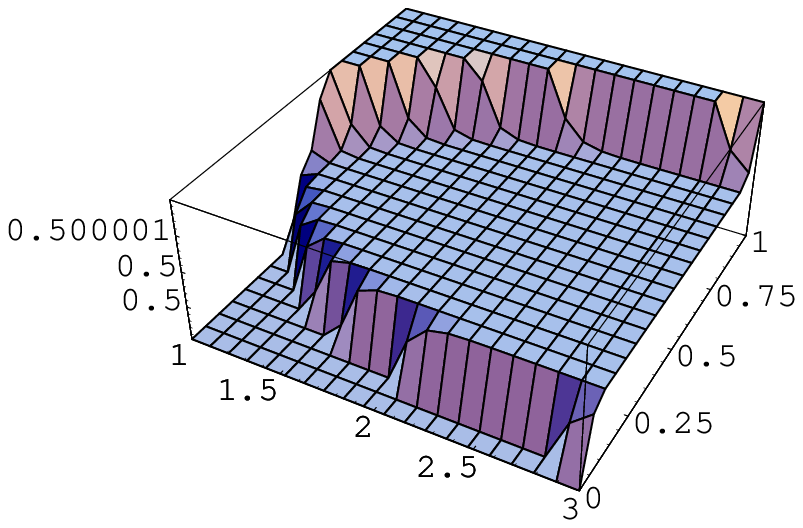}
\put(-20,40){$\eta$} \put(-250,40){$\eta$} \put(-150,15){$\alpha$}
\put(-370,15){$\alpha$} \put(-400,120){$(a)$}
\put(-175,120){$(b)$} \put(-450,70){$\mathcal{F}$}
 \put(-200,65){$\mathcal{F}$}
\end{center}
\caption{The fidelity of the teleported state for  (a) $m=3$,
(b)$m=5$.}
\end{figure}
We can summarize the preceding result as: It is possible to
teleport a multipartite ECS with high degree of fidelity and
efficiency. The mean photon number $\alpha$ plays the central role
on controlling  the efficiency of transporting  multipartite ECS.
For small  values of the mean photon number the effect of the
noise channel appears clearly and it is considered the controller
parameter

\section{Conclusion}

In conclusion, we have proposed a general quantum teleportation protocol to
teleport multipartite of entangled coherent states. In this scheme, one can
generate the multipartite quantum channels by using a series of beam
splitters and phase shifters. Also, we describe the theoretical technique to
generate a multipartite quantum channels. It is shown that the probability
of successes is $0.5$ does not depend on the channel parameters. One of the
most advantage of this protocol is, it works with the same efficiency even
the modes are even or odd numbers. So it can teleport all class of
multipartite coherent states with high efficiency. The possibility, of
applying this protocol in the presences of noise quantum channels is
investigated, where we consider the noise due to the photons absorption
losses. The noise strength plays the central role on the fidelity of the
teleported state for small value of the channel parameter $\alpha$. However,
for larger value of $\alpha$, the fidelity decreases very fast reaches it
minimum value, $0.5$.

\end{document}